\documentstyle[sprocl]{article}

\input{psfig}

\def\VEV#1{\left\langle #1\right\rangle}
\def\sec{\ifmmode \,\, {\rm sec} \else sec \fi}
\def\eV {\ifmmode \,\, {\rm eV} \else eV \fi}
\def\keV{\ifmmode \,\, {\rm keV} \else keV \fi}
\def\MeV{\ifmmode \,\, {\rm MeV} \else MeV \fi}
\def\GeV{\ifmmode \,\, {\rm GeV} \else GeV \fi}
\def\TeV{\ifmmode \,\, {\rm TeV} \else TeV \fi}
\def\fm{\ifmmode \,\, {\rm fm} \else TeV \fi}
\def\pbarn{\ifmmode \,\, {\rm pb} \else pb \fi}
\def\km{\ifmmode {\rm km}\, \else km \fi}
\def\Mpc{\ifmmode {\rm Mpc}\, \else Mpc \fi}
\def\Gyr{\ifmmode {\rm Gyr}\, \else Gyr \fi}

\def\fun#1#2{\lower3.6pt\vbox{\baselineskip0pt\lineskip.9pt
  \ialign{$\mathsurround=0pt#1\hfil##\hfil$\crcr#2\crcr\sim\crcr}}}
\def\la{\mathrel{\mathpalette\fun <}}
\def\ga{\mathrel{\mathpalette\fun >}}

\def\sbar#1{\kern 0.8pt
        \overline{\kern -0.8pt #1 \kern -0.8pt}
        \kern 0.8pt}  

\def\meter{\ifmmode \,\, {\rm m} \else m \fi}
\def\yr {\ifmmode \,\, {\rm yr} \else yr \fi}

\def\sr{\ifmmode \,\, {\rm sr} \else sr \fi}

\def\hatn{{\bf \hat n}}

\def\be{\begin{equation}}
\def\ee{\end{equation}}
\def\bea{\begin{eqnarray}}
\def\eea{\end{eqnarray}}

\begin{document}


\title{NEW TESTS OF INFLATION\footnote{ To appear in {\it
PASCOS-98}, proceedings of the conference, Boston, MA, March
22--29, 1998, edited by P. Nath (World Scientific, Singapore).}}

\author{M. KAMIONKOWSKI}

\address{Department of Physics \\ Columbia University \\ 538
West 120th Street \\ New York, NY~~USA \\E-mail: kamion@phys.columbia.edu}

\maketitle\abstracts{
Slow-roll inflation generically makes several predictions: a
flat Universe, primordial adiabatic density perturbations, and a
stochastic gravity-wave background. Each inflation model will
further predict specific relations between the amplitudes and
shapes of the spectrum of density perturbations and gravity
waves.  There are now excellent prospects for testing precisely
these predictions with forthcoming cosmic microwave background
(CMB) temperature and polarization maps.}

\section{Introduction}

Although the physics responsible for slow-roll inflation is
still not well understood,
inflation generically predicts (1) a flat Universe; (2) that
primordial adiabatic (scalar metric) perturbations are
responsible for the large-scale structure (LSS) in the Universe
today~\cite{scalars}; and (3) a stochastic gravity-wave
background (tensor metric
perturbations)~\cite{abbott}. Furthermore, each inflationary
model predicts (4) specific relations between the
``inflationary observables,'' the amplitudes  and
spectral indices of the scalar and tensor perturbations
\cite{steinhardt}.
The amplitude of the gravity-wave background is
proportional to the height of the inflaton potential.
Therefore, the height of the inflaton potential, $V(\phi)$, can
be fixed by the tensor contribution to the CMB quadrupole
moment, $C_2^{\rm TT}$:
\begin{equation}
     {\cal T} \equiv  6\, C_2^{{\rm TT},{\rm tensor}}= 9.2
     \,V/m_{\rm Pl}^4.
\label{amplitudes}
\end{equation}
The predictions for the scalar amplitude and the spectral
indices follow immediately from the shape of the inflaton
potential.  Therefore, determination of the inflationary
observables would illuminate the physics responsible for
inflation.

Until recently, none of these predictions could be tested with
precision.  Measured values for the density of the Universe span
almost an order of magnitude.  Furthermore, most measurements do not probe
the possible contribution of a cosmological constant (or some
other diffuse matter component), so they do not address the
geometry of the Universe.  The only observable effects of a
stochastic gravity-wave background are in the CMB.  COBE
observations do in fact provide an upper limit to the tensor
amplitude, and therefore an inflaton potential, $(V/m_{\rm
Pl}^4)\la 5\times 10^{-12}$.  However, there is no way to
disentangle the scalar and tensor contributions to the COBE
anisotropy.

In recent years, it has become increasingly likely that
adiabatic perturbations are responsible for the
origin of structure.  Before COBE, there were numerous plausible
models for structure formation: e.g., isocurvature perturbations
both with and without cold dark matter, late-time or slow phase 
transitions, topological defects (cosmic strings or textures),
superconducting cosmic strings, explosive or seed models, a
``loitering'' Universe, etc.  However, the amplitude of the COBE 
anisotropy makes all these alternative models unlikely.  With
adiabatic perturbations, hotter regions at the surface of last
scatter are embedded in deeper potential wells, so the
reddening due to the the gravitational redshift of the photons
from these regions partially cancels the higher intrinsic
temperatures.  Thus, other models will generically produce more
anisotropy for the same density perturbation.  When
normalized to the density fluctuations indicated by galaxy
surveys, alternative models thus generically produce a larger
temperature fluctuation than that measured by COBE~\cite{jaffe}.
In the past year, some leading proponents of topological defects
have conceded that these models have difficulty accounting for
the origin of large-scale structure~\cite{towel}.

We are now entering an exciting new era, driven by new
theoretical ideas and developments in detector technology, in
which the predictions of inflation will be tested with
unprecedented precision.  It is even conceivable that early in
the next century, we will move from verification of inflation to
direct investigation of the high-energy physics responsible for
inflation.

The purpose of this talk is to review how forthcoming CMB
experiments will test several of these predictions.
I will first discuss how CMB temperature anisotropies will test
the inflationary predictions of a flat Universe and a primordial
spectrum of density perturbations.  I will then review how a CMB
polarization map may be used to isolate the gravity waves and
briefly review how detection of these tensor modes may be used
to learn about the physics responsible for inflation.  

\section{Temperature Anisotropies}

The primary goal of CMB experiments that map the temperature as
a function of position on the sky is recovery of the
temperature autocorrelation function or angular power spectrum
of the CMB.  The fractional temperature perturbation
$\Delta T(\hatn)/T$ in a given direction $\hatn$ can be expanded
in spherical harmonics,
\begin{equation}
     {\Delta T(\hatn) \over T} = \sum_{lm} \, a_{(lm)}^{\rm T}\,
     Y_{(lm)}(\hatn),
\label{eq:Texpansion}
\end{equation}
where the multipole coefficients are given by
\begin{equation}
     a_{(lm)}^{\rm T} = \int\, d\hatn\, Y_{(lm)}^*(\hatn) \, {\Delta
     T(\hatn) \over T}.
\label{eq:alms}
\end{equation}
Statistical isotropy and homogeneity of the Universe imply that
these coefficients have expectation values $\VEV{ (a_{(lm)}^{\rm
T})^*
a_{(l'm')}^{\rm T}} = C_l^{\rm TT} \delta_{ll'} \delta_{mm'}$ when
averaged over the sky.  Roughly speaking, the multipole moments
$C_l^{\rm TT}$ measure the mean-square temperature difference
between two points separated by an angle $(\theta/1^\circ) \sim
200/l$.

\begin{figure*}[htbp]
\centerline{\psfig{file=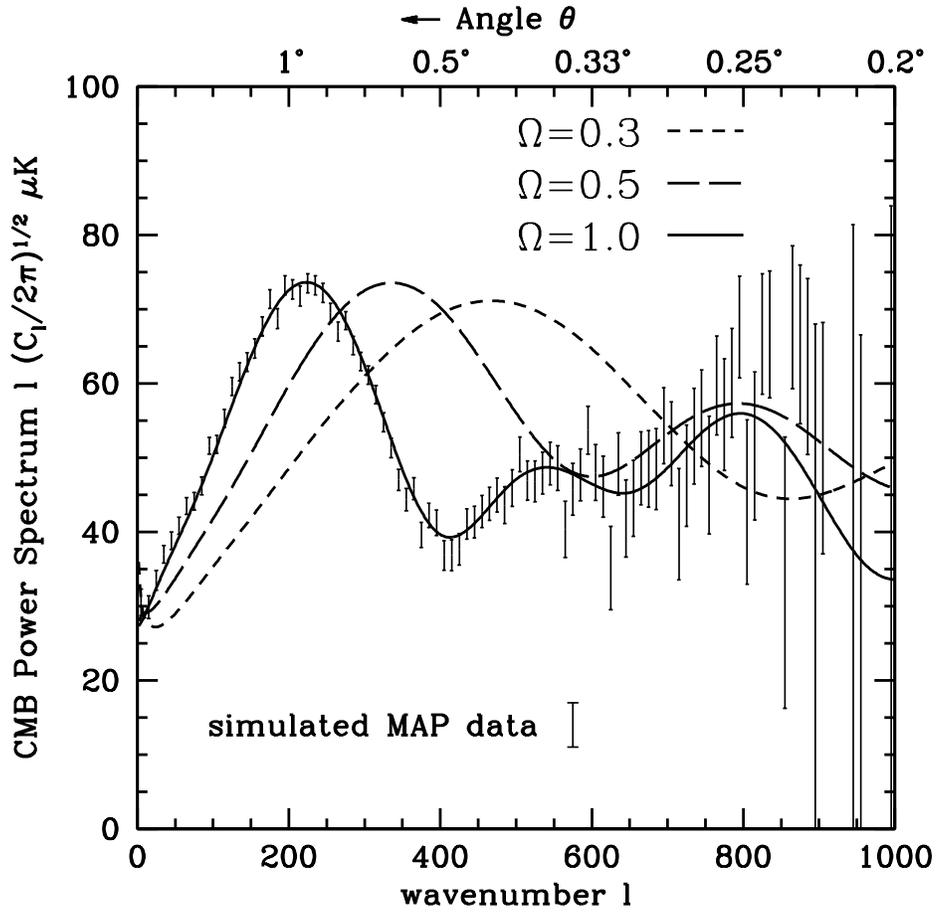,width=5in}}
\bigskip
\caption{
	  Theoretical predictions and current and simulated data 
	  for CMB spectra as a function of multipole moment $l$
	  for models with primordial adiabatic perturbations.
	  The curves are for three different values of the total 
	  density $\Omega$.  Simulated MAP data points are
	  shown.}
\label{fig:models}
\end{figure*}

Predictions for the $C_l$'s can be made given a theory for
structure formation and the values of several cosmological
parameters.  The curves in Fig. \ref{fig:models} show these
predictions for several value of $\Omega$.~\cite{science}\ \   The
bumps come from
oscillations in the photon-baryon fluid at the surface of last
scatter.  As the Figure shows, these small-angle
CMB anisotropies can be used to determine the geometry of the
Universe~\cite{kamspergelsug}.  The angle
subtended by the horizon at the surface of last scatter is
$\theta_H \sim \Omega^{1/2} \;1^\circ$, and the peaks in the CMB
spectrum are due to causal processes at the surface of last
scatter.  Therefore, the angles (or values of $l$) at which the
peaks occur determine the geometry of the Universe.  Detailed
calculations also show that the angular position of the first  peak is
relatively insensitive to the values of other undetermined (or
still imprecisely determined) cosmological parameters such as
the baryon density, the Hubble constant, and the cosmological
constant (as well as several others).  Therefore,
determination of the location of this first acoustic peak should
provide a robust measure of the geometry of the Universe.

The precision attainable is ultimately limited by
cosmic variance and practically by the finite angular resolution,
instrumental noise, and partial sky coverage in a realistic CMB
mapping experiment.  Taking these considerations into account,
it can be shown that future satellite missions may
potentially determine $\Omega$ to better than 10\% {\it after}
marginalizing over all other undetermined parameters, and better
than 1\% if the other parameters can be fixed by independent
observations or assumption~\cite{jkksone}.  This would be far
more accurate than any traditional determinations of the geometry.

It can similarly be shown that the CMB should provide determinations of
the cosmological constant and baryon density far more precise
than those from traditional observations~\cite{jkkstwo}.  If there is more
nonrelativistic matter in the Universe than baryons can account
for---as suggested by current observations---it will become
increasingly clear with future CMB measurements.  Subsequent
analyses have confirmed these estimates with more precise
numerical calculations~\cite{bet}.

Although these forecasts relied on the assumptions that
adiabatic perturbations were responsible for structure formation
and that reionization would not erase CMB anisotropies, these
assumptions have become increasingly
justifiable in the past few years.  As discussed above,
the leading alternative theories for structure formation now
appear to be in trouble, and recent detections of CMB
anisotropy at degree angular separations show that the effects
of reionization are small.  

NASA has recently approved the flight of a satellite mission,
the Microwave Anisotropy Probe (MAP)~\cite{MAP}, in the year 2000
to carry out these measurements, and ESA has approved the
flight of a subsequent more precise experiment, the Planck
Surveyor~\cite{PLANCK}.  Therefore, it appears increasingly
likely that the inflationary prediction of a flat Universe will
be carried out precisely in the near future.

The predictions of a nearly scale-free spectrum of primordial
adiabatic perturbations will also be further tested with
measurements of small-angle CMB anisotropies.  The existence and
structure of the acoustic peaks
will provide an unmistakable signature of adiabatic
perturbations~\cite{huwhite} and the spectral index $n_s$ can be determined
from fitting the theoretical curves to the data in the same way
that the density, cosmological constant, baryon density, and
Hubble constant are also fit~\cite{jkkstwo}.

Temperature anisotropies produced by a stochastic gravity-wave
background would affect the shape of the angular CMB spectrum,
but there is no way to disentangle the scalar and tensor
contributions to the CMB anisotropy in a model-independent way.
Unless the tensor signal is large, the cosmic variance from the
dominant scalar modes will provide an irreducible limit to the
sensitivity of a temperature map to a tensor signal~\cite{jkkstwo}.

\section{CMB Polarization and Gravitational Waves}

Although a CMB temperature map cannot unambiguously distinguish
between the density-perturbation and gravity-wave contributions
to the CMB, the two can be decomposed in a model-independent
fashion with a map of the CMB polarization
\cite{probe,ourpolarization,selzald}.  Suppose we
measure the linear-polarization ``vector'' $\vec P(\hatn)$ at
every point $\hatn$ on the sky.\footnote{Strictly
speaking, the linear polarization does not transform as a
vector, but the argument given here generalizes when one
describes the polarization state properly as a symmetric trace-free
$2\times2$ tensor.}  Such a vector field can be written as the
gradient of a scalar function $A$ plus the curl of a vector
field $\vec B$,
\begin{equation}
     \vec P(\hatn) \, = \, \vec \nabla A \, + \, \vec\nabla \times \vec
     B.
\label{eq:curl}
\end{equation}
The gradient (i.e., curl-free) and curl components can be
decomposed by taking the divergence or curl of $\vec
P(\hatn)$ respectively.  Density perturbations are scalar metric
perturbations, so they have no handedness.  They can therefore
produce no curl.  On the other hand, gravitational waves {\it
do} have a handedness so they can (and we have shown that they
do) produce a curl.  This therefore provides a way to detect the
inflationary stochastic gravity-wave background and thereby
test the relations between the inflationary observables.  It
should also allow one to determine (or at least constrain in the
case of a nondetection) the height of the inflaton potential.

As with a temperature map, the sensitivity of a polarization map
to gravity waves will be determined by the
instrumental noise and fraction of sky covered, and by the
angular resolution.  Suppose the detector sensitivity is $s$ and
the experiment lasts for $t_{\rm yr}$ years with an angular
resolution better than $1^\circ$.  Suppose further that we
consider only the curl component of the polarization in our
analysis.  Then the smallest tensor amplitude ${\cal T}_{\rm
min}$ to which the experiment will be sensitive at $1\sigma$ is
\cite{detectability}
\begin{equation}
     {{\cal T}_{\rm min}\over 6\, C_2^{\rm TT}}
      \simeq 5\times 10^{-4} \left( {s\over \mu{\rm K}\,\sqrt{\rm
      sec}} \right)^2 t_{\rm yr}^{-1}.
\label{CCresult}
\end{equation}
Thus, the curl component of a full-sky polarization map is
sensitive to inflaton potentials $(V/m_{\rm Pl}^4)\ga 5 \times
10^{-15}t_{\rm yr}^{-1}$ $(s/\mu{\rm K}\, \sqrt{\rm sec})^2$.  
Improvement on current constraints with only the curl
polarization component requires a detector sensitivity
$s\la40\,t_{\rm yr}^{1/2}\,\mu$K$\sqrt{\rm sec}$.  For
comparison, the detector sensitivity of MAP will be $s={\cal
O}(100\,\mu$K$\sqrt{\rm sec})$.  However, Planck may conceivably
get sensitivities around $s=25\,\mu$K$\sqrt{\rm sec}$.

Even a small amount of reionization will significantly increase
the polarization signal at low $l$~\cite{reionization}.
For example, suppose the optical depth to the surface of last
scatter is $\tau=0.1$.  With such a level of
reionization, the sensitivity to the tensor amplitude is
increased by more than a factor of 5 over that in
Eq. (\ref{CCresult}).  This level of reionization (if not more)
is expected in cold-dark-matter models
\cite{kamspergelsug,blanchard,haiman}, so if anything,
Eq.~(\ref{CCresult}) provides a conservative estimate.

Furthermore, the estimate in Eq. (\ref{CCresult}) takes into
account only the information provided by the curl polarization
moments.  A complete likelihood analysis will fit the
temperature-polarization map to the temperature moments, the
gradient component, and the temperature-polarization
cross-correlation, and this will improve the sensitivity
significantly over that given in Eq. (\ref{CCresult})
\cite{detectability}.

\section{Discussion}

If MAP and Planck find a CMB temperature-anisotropy spectrum
consistent with a flat Universe and nearly--scale-free
primordial adiabatic perturbations, then the next step will be
to isolate the gravity waves with the polarization of the CMB.
If inflation has something to do with grand unification, then it
is possible that Planck's polarization sensitivity will be
sufficient to see the polarization signature of gravity waves.
However, it is also quite plausible that the height of the
inflaton potential may be low enough to elude detection by
Planck.  If so, then a subsequent experiment with better
sensitivity to polarization will need to be done.

Inflation also predicts that the distribution of primordial
density perturbations is gaussian, and this can be tested with
CMB temperature maps and with the study of the large-scale
distribution of galaxies.  Since big-bang nucleosynthesis
predicts that the baryon density is $\Omega_b \la 0.1$ and
inflation predicts $\Omega=1$, another prediction of inflation
is a significant component of nonbaryonic dark matter.  This can
be either in the form of vacuum energy (i.e., a cosmological
constant), and/or some new elementary particle.  Therefore,
discovery of particle dark matter could be interpreted as
evidence for inflation.

\section*{Acknowledgments}
This work was supported by D.O.E. contract DEFG02-92-ER 40699,
NASA NAG5-3091, and the Alfred P. Sloan Foundation.

\end{document}